\documentclass{PoS}

\title{Cusps in $\eta'\to\eta\pi\pi$ decays}

\ShortTitle{Cusps in $\eta'\to\eta\pi\pi$ decays}

\author{\speaker{Sebastian P. Schneider}\\
        Helmholtz-Institut f\"ur Strahlen- und Kernphysik (Theorie)
   and 
   Bethe Center for Theoretical Physics,
   Universit\"at Bonn, D--53115~Bonn, Germany\\
        E-mail: \email{schneider@hiskp.uni-bonn.de}}

\author{Bastian Kubis\\
        Helmholtz-Institut f\"ur Strahlen- und Kernphysik (Theorie)
   and 
   Bethe Center for Theoretical Physics,
   Universit\"at Bonn, D--53115~Bonn, Germany\\
        E-mail: \email{kubis@hiskp.uni-bonn.de}}

\abstract{The discovery of the cusp effect in the decay $K^+\to\pi^+\pi^0\pi^0$ has spurred the search for other decay channels, where this phenomenon, which is generated by strong final-state interactions, should also occur. A very promising candidate is $\eta'\to\eta\pi^0\pi^0$. The cusp effect offers an excellent opportunity to experimentally extract $\pi\pi$ S-Wave scattering lengths. We adapt and generalize the non-relativistic effective field theory framework developed for $K\to3\pi$ decays to $\eta'\to\eta\pi\pi$. The cusp effect is predicted to have an effect of more than 8\% on the decay spectrum below the $\pi^+\pi^-$ threshold \cite{BKSPS}. We also show that with our current theoretical information about $\eta'\to\eta\pi\pi$ decays, it is not possible to extract $\pi\eta$ threshold parameters.}

\FullConference{6th International Workshop on Chiral Dynamics, CD09\\
		July 6-10, 2009\\
		Bern, Switzerland}

\begin{document}

\section{Introduction}

The investigation of the cusp effect in the decay $K^+\to\pi^0\pi^0\pi^+$ is one of the most precise methods to extract S-wave $\pi\pi$ scattering lengths from experiment~\cite{Cabibbo,CI,CGKR,NA48}. A heuristic explanation for the cusp effect is as follows: the cusp in the invariant mass spectrum of the $\pi^0\pi^0$ pair is generated by the decay $K^+\to\pi^+\pi^+\pi^-$ followed by charge-exchange rescattering $\pi^+\pi^-\to\pi^0\pi^0$, plus the fact that the pion mass difference shifts the $\pi^+\pi^-$ threshold into the physical region. The large branching ratio of $K^+\to\pi^+\pi^+\pi^-$ compared to $K^+\to\pi^0\pi^0\pi^+$ generates a sizable perturbation of the decay spectrum of the latter, so that this channel is well-suited for a cusp analysis.

In this respect, $\eta'\to\eta\pi^0\pi^0$ decays, where in the isospin limit one finds $\textrm{BR}(\eta'\to\eta\pi^+\pi^-) = 2 \,\textrm{BR}(\eta'\to\eta\pi^0\pi^0)$, offers a very promising candidate for an alternative study of the cusp. This study is very timely with regards to the upcoming high-statistics $\eta'$ experiments at ELSA, MAMI-C, WASA-at-COSY, KLOE-at-DA$\Phi$NE, or BES-III, which should increase the data basis on $\eta'$ decays by orders of magnitude.

\section{The decay amplitude for $\eta'\to\eta\pi\pi$}
We have used the modified non-relativistic effective field theory framework~\cite{CGKR, Photons} to calculate the $\eta'\to\eta\pi\pi$ decay amplitude, which -- while leading to manifestly covariant results -- produces the correct analytic structure of the decay amplitude in the low-energy region. This non-relativistic effective field theory exhibits a consistent power counting scheme: We introduce a formal parameter $\epsilon$ and set up a set of rules by counting three-momenta as $\mathcal{O}(\epsilon)$, kinetic energies as $T_i=p_i^0-M_i$ as $\mathcal{O}(\epsilon^2)$, masses as $\mathcal{O}(1)$ and the $Q$-value $M_\eta'-M_{\eta}-2M_\pi=\sum_iT_i$ as $\mathcal{O}(\epsilon^2)$.

Performing a correlated expansion in terms of $\epsilon$ and the small scattering lengths of $\pi\pi$ and $\pi\eta$ scattering, $a_{\pi\pi}$, $a_{\pi\eta}$ or generically $a$, we have calculated the decay amplitude up-to-and-including terms of $\mathcal{O}(\epsilon^4)$ at tree-level, $\mathcal{O}(a\epsilon^5)$ at one-loop level and $\mathcal{O}(a_{\pi\pi}^2\epsilon^6,a_{\pi\pi}a_{\pi\eta}\epsilon^2,a_{\pi\eta}^2\epsilon^2)$ at two-loop level. The amplitudes are ready for use in experiments and can be found in~\cite{BKSPS}, a FORTRAN code of the amplitude is available upon request. Notice in particular the high accuracy we have achieved in terms of $a_{\pi\pi}$. The amplitude obtained should thus allow for a precise extraction of the $\pi\pi$ scattering lengths. Radiative corrections have been taken into account consistently at $\mathcal{O}(e^2a_{\pi\pi}\log\epsilon)$ in $\eta'\to\eta\pi^0\pi^0$ and at $\mathcal{O}(e^2a^0)$ in the auxiliary $\eta'\to\eta\pi^+\pi^-$ channel, based on the findings in~\cite{Photons}. We have also addressed the issue of six-particle rescattering vertices and inelastic channels and shown that their contributions are sufficiently small to be neglected.

\section{Prediction of the cusp}

Since there is no precision data on $\eta'\to\eta\pi\pi$ decays yet, we predict the expected size and shape of the cusp effect, instead of extracting the $\pi\pi$ threshold parameters from data. We use the theoretical values for the $\pi\pi$ threshold parameters $a_0=0.220 \pm 0.005$, $a_2 = -0.0444\pm 0.0010$, $b_0 = (0.276 \pm 0.006)\times M_\pi^{-2}$, $b_2= (-0.0803 \pm 0.0012)\times M_\pi^{-2}$ as input parameters~\cite{CGL}.

There is no experimental data on $\pi\eta$ threshold parameters, yet. We therefore perform a low-energy expansion of the one-loop chiral perturbation theory amplitude~\cite{BKM:pieta} to extract these. We find, however, that $\pi\eta$ threshold parameters are very badly constrained by chiral perturbation theory. For our analysis we vary these parameters according to $\bar a_0 = (0 \ldots +16)\times 10^{-3}$ and $\bar b_0 = (0 \ldots +10) \times 10^{-3} M_\pi^{-2}$.
Due to the smallness of the $\pi\eta$ P-wave $\bar a_1$ we neglect its contribution.

We will use the central values (without errors) of the most recent determinations of the $\eta'\to\eta\pi\pi$ Dalitz plot by the VES collaboration for the charged channel~\cite{VES}, which, adjusting the normalization to the neutral decay, read $a=-0.133$, $b=-0.116$, $d=-0.094$. The charged and the neutral decay amplitude are related by isospin symmetry in the tree-level couplings. We have re\-adjusted the input parameters for the low-energy couplings of the Dalitz plot in such a way that the {\it full} amplitude squared expanded in terms of the Dalitz variables $x$ and $y$ reproduces the VES parameters. Performing such a renormalization procedure for each set of $\pi\eta$ parameters we observe that the variations due to $\pi\eta$ threshold parameters can be reabsorbed in the low-energy couplings of the polynomial expansion.
\begin{figure}
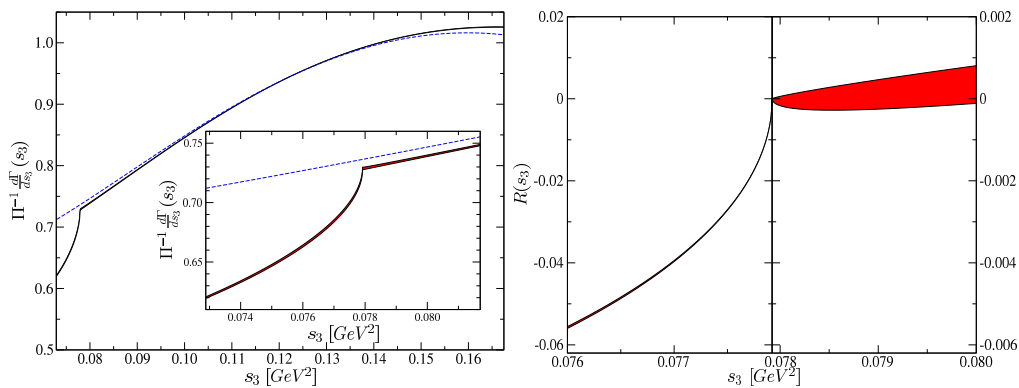

 \centering
 \includegraphics[height=5cm]{REN_color.eps}
 \includegraphics[height=5cm]{CUSP_color.eps}
 \caption{Left: the differential decay rate ${d\Gamma}/{ds_3}$ divided by phase space for the tree (dashed blue) and the full (red band) amplitude. The cusp effect below the charged pion threshold is prominently visible in the invariant $\pi^0\pi^0$ spectrum. Right: difference between full and tree decay rate ${d\Gamma}/{ds_3}$ in the cusp region, divided by the phase space, and shifted to 0 at $s_3=4M_\pi^2$. Note that the scale above the cusp has been increased by a factor of 10.}
 \label{fig:Twoloopwren}
\end{figure}

The differential decay rate as a function of invariant mass of the $\pi^0\pi^0$ pair is displayed in Fig.~\ref{fig:Twoloopwren}. The cusp below the charged pion threshold gives rise to an integrated event deficit of $\sim 8\%$, which is comparable in size to $13\%$ in $K^+\to\pi^+\pi^0\pi^0$. We also show a zoom into the cusp region, where we plot the decay spectrum with the tree spectrum subtracted and shifted to $0$ at threshold,

The two-loop cusp is found to be strongly suppressed. The perturbation of the spectrum above the threshold amounts to $\sim 0.5\%$. The explanation for this is found resorting to the threshold theorem and is due to the approximate isospin symmetry between charged and neutral decay channel~\cite{BKSPS}. The threshold theorem also allows for an estimate of three-loop contributions. We find a reduction of the cusp by $\sim 0.5\%$ due to three-loop effects. The decay amplitude is thus completely dominated by one-loop effects.

\section{On the extraction of $\pi\eta$ threshold parameters}
\begin{figure}
 \centering
 \includegraphics[height=2.4cm]{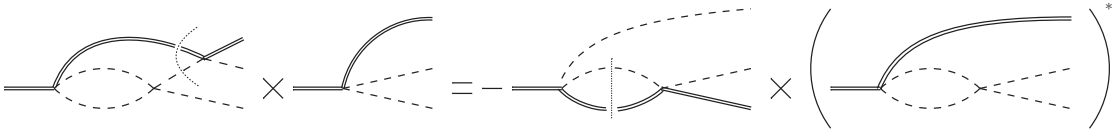}
 \caption{Visualization of the exact cancellation of the $\pi\eta$ two-loop cusp at $s_1=(M_\pi+M_\eta)^2$.}
 \label{fig:pietatwoloopcanc}
\end{figure}

In principle the $\eta'\to\eta\pi\pi$ decay channel should offer access to $\pi\eta$ threshold parameters. Since the opening of the $\pi\eta$ channel is at the border of the Dalitz plot there is no one-loop cusp in the physical region as in $\pi\pi$ scattering. One might still wonder about possible two-loop effects above threshold. Casting aside experimental difficulties arising from precision studies at the border of the Dalitz plot, we address this issue and attempt to quantify the effect. We find, however, that there is an {\it exact} cancellation between the product of two-loop and tree graph and the product of two one-loop graphs as illustrated in Fig.~\ref{fig:pietatwoloopcanc}. Without exact theoretical knowledge of the tree-level couplings an extraction of $\pi\eta$ threshold parameters in this framework is not possible.

\section*{Acknowledgments}
Partial financial support by the Helmholtz Association through funds provided
to the virtual institute ``Spin and strong QCD'' (VH-VI-231), 
by the European Community-Research Infrastructure Integrating Activity 
``Study of Strongly Interacting Matter''
(acronym HadronPhysics2, Grant Agreement n.~227431) under the Seventh 
Framework Programme of the EU, by DFG (SFB/TR 16, ``Subnuclear Structure of Matter'') and by the Bonn-Cologne Graduate School of Physics and Astronomy is gratefully
acknowledged.


\begin{thebibliography}{99}
\bibitem{BKSPS}
  B.~Kubis and S.~P.~Schneider,
  \emph{The cusp effect in eta' --> eta pi pi decays},
  \emph{Eur.\ Phys.\ J.\  C} {\bf 62} (2009) 511
  [{\tt 0904.1320 [hep-ph]}].
  %%CITATION = EPHJA,C62,511;%%

\bibitem{Cabibbo}
  N.~Cabibbo,
  \emph{Determination of the a0-a2 pion scattering length from K->pi+ pi0 pi0 decay},
  \emph{Phys.\ Rev.\ Lett.\ }  {\bf 93} (2004) 121801
  [{\tt hep-ph/0405001}].
  %%CITATION = PRLTA,93,121801;%%

\bibitem{CI}
  N.~Cabibbo and G.~Isidori,
  \emph{Pion-pion scattering and the K->3pi decay amplitudes},
  \emph{JHEP} {\bf 0503} (2005) 021
  [{\tt hep-ph/0502130}].
  %%CITATION = JHEPA,0503,021;%%

\bibitem{CGKR}
  G.~Colangelo, J.~Gasser, B.~Kubis and A.~Rusetsky,
  \emph{Cusps in K --> 3pi decays},
  \emph{Phys.\ Lett.\ B} {\bf 638} (2006) 187
  [{\tt hep-ph/0604084}].
  %%CITATION = PHLTA,B638,187;%%

\bibitem{NA48}
 J.~R.~Batley {\it et al.}  [NA48/2 Collaboration],
  \emph{Observation of a cusp-like structure in the pi0 pi0 invariant mass
  distribution from K+- --> pi+- pi0 pi0 decay and determination of the  pi pi
  scattering lengths},
  \emph{Phys.\ Lett.\  B} {\bf 633} (2006) 173
  [{\tt hep-ex/0511056}].
  %%CITATION = PHLTA,B633,173;%%

\bibitem{Photons}
  M.~Bissegger, A.~Fuhrer, J.~Gasser, B.~Kubis and A.~Rusetsky,
  \emph{Radiative corrections in K --> 3 pi decays},
  \emph{Nucl.\ Phys.\  B} {\bf 806} (2009) 178
  [{\tt 0807.0515 [hep-ph]}].
  %%CITATION = NUPHA,B806,178;%%

\bibitem{CGL}
  G.~Colangelo, J.~Gasser and H.~Leutwyler,
  \emph{$\pi\pi$ scattering},
  \emph{Nucl.\ Phys.\  B} {\bf 603} (2001) 125
  [{\tt hep-ph/0103088}].
  %%CITATION = NUPHA,B603,125;%%

\bibitem{BKM:pieta}
  V.~Bernard, N.~Kaiser and U.-G.~Mei{\ss}ner,
  \emph{pi eta scattering in QCD},
  \emph{Phys.\ Rev.\  D} {\bf 44} (1991) 3698.
  %%CITATION = PHRVA,D44,3698;%%

\bibitem{VES}
  V.~Dorofeev {\it et al.},
  \emph{Study of eta' --> eta pi+ pi- Dalitz plot},
  \emph{Phys.\ Lett.\  B} {\bf 651} (2007) 22
  [{\tt hep-ph/0607044}].
  %%CITATION = PHLTA,B651,22;%%

\end{thebibliography}
\end{document}